\documentclass[pre,twocolumn,superscriptaddress,showpacs]{revtex4}
\usepackage{graphicx,color}
\usepackage{amsmath}
\definecolor{r}{rgb}{1,0,0}

\begin{document}

\newcommand{\LANL}{Condensed Matter and Thermal Physics and Center for Nonlinear Studies, 
Los Alamos National Lab, NM, 87545, USA}

\newcommand{\SZFKI}{Research Institute for Solid State Physics and Optics,
                    POB 49, H-1525 Budapest, Hungary}

\title{Rapid granular flows on a rough incline: phase diagram, gas transition, and effects of air drag}

\author{Tam\'as B\"orzs\"onyi}
\email{btamas@szfki.hu}
  \affiliation{\LANL}
  \affiliation{\SZFKI}

\author{Robert E. Ecke}
  \affiliation{\LANL}

\begin{abstract} 
We report experiments on the overall phase diagram of granular flows on an incline 
with emphasis on high inclination angles where the mean layer velocity approaches 
the terminal velocity of a single particle free falling in air. The granular flow 
was characterized by measurements of the surface velocity, the average layer height, 
and the mean density of the layer as functions of the hopper opening, the plane 
inclination angle and the downstream distance $x$ of the flow. At high inclination 
angles the flow does not reach an $x-$invariant steady state over the 
length of the inclined plane. 
For low volume flow rates, a transition was detected between dense and 
very dilute (gas) flow 
regimes.  We show using a vacuum flow channel that air did not qualitatively change 
the phase diagram and did not quantitatively modify mean flow velocities of the 
granular layer except for small changes in the very dilute gas-like phase.
\end{abstract}


\maketitle

\section{Introduction}
\label{intro}

Granular flows are often separated into two categories dealing with 
(i) dense granular flows where the flow density is not far from 
the density of static packing (for studies in 2 dimensions see 
\cite{an2001,ra2003,bede2003,ra2004,ra2005} or in  3 dimensions see 
\cite{humo1984,po1999,fopo2001,sa1979,fopo2003,hawa2000,gdrmidi2004}) 
and (ii) a rapid, very dilute (gas) regime \cite{azch1999}, which is often 
analyzed in terms of a kinetic theory \cite{go2003}).  Although flow on a rough 
inclined plane has become a model experiment because of its simple geometry most 
of the available research focuses on the dense regime while less data is devoted 
to characterizing the gas regime and the transition between the dense and very 
dilute phase regimes. 

A granular layer of thickness $h$ on a rough inclined plane starts flowing only 
if the plane inclination $\theta$ surpasses a critical angle $\theta_c$ and stops 
flowing when $\theta$ is decreased below an angle of repose $\theta_r$
(for reviews see: \cite{jana1996,gdrmidi2004,arts2006}). 
Experiments in 3 dimensions typically feature a storage container with an opening of height 
$H$ that influences the volume flow rate of material down the plane 
\cite{po1999,fopo2001,sa1979,fopo2003,hawa2000}. 
The properties of granular flows in this 
system have been extensively studied when $\theta$ is not much larger than 
the flow initiation conditions \cite{po1999,sa1979,fopo2003,hawa2000}. 
For $\theta \approx \theta_r$ and for small volume flow rates, intermittent 
flows are observed \cite{si2005} including avalanches \cite{da2000} 
and wave-like motion \cite{sa1979,fopo2003,hawa2000}.
At somewhat larger $\theta$, the grains flow uniformly with a statistically steady 
flow velocity that depends on the layer height \cite{po1999,fopo2003,hawa2000}. 
For still 
higher $\theta$, the flow properties have not been well studied although an 
interesting stripe state has been reported \cite{fopo2001,fopo2002,boec2005}. 
One might expect that the layer density would decrease as $\theta$ is increased 
because of higher average flow velocity which creates larger shear rates and higher 
granular temperature. Because the ratio of the down-plane force to the normal 
force diverges at $\theta = \pi/2$, one might also ask whether a dense phase 
with a well defined layer thickness continues to exist at large $\theta$.

Flowing granular layers can be described by a set of macroscopic variables
that depend on the system control parameters. The two control parameters for
granular flow on a rough inclined plane are the hopper opening $H$ and the 
inclination angle $\theta$. The hopper opening largely influences the volume
flow rate from the hopper whereas $\theta$ controls the balance of tangential 
and normal gravitational force on the layer. An additional parameter that is 
harder to vary systematically is the roughness of the inclined plane surface.
For fixed surface roughness, the flow properties of the layer at fixed $H$ 
and $\theta$ can be characterized by the surface velocity, the height $h$ 
and the average density $\rho$. These quantities depend on the control
parameters and on each other in a complicated manner. Part of our purpose 
in this paper is to understand these relationships. 

One particular issue for granular flows at high inclination angles was suggested
by idealized numerical simulations in which gravitational forcing cannot be 
balanced by energy dissipation mechanisms, and the flow is predicted to accelerate 
\cite{sier2001,sila2003}. 
Constitutive equations recently proposed for 
dense granular flows suggest that the effective friction coefficient saturates to 
a finite value $\mu_2$ for high shear rates \cite{po1999,jofo2005,jofo2006}. 
In this case the flow is expected to accelerate for $\theta$ above 
$\theta_2=$atan$(\mu_2)$.
As we will show in the present work, in accordance with the data presented in
\cite{po1999} the dense, non-accelerating regime can only be observed up to
tan$\theta/$tan$\theta_2=0.85$.
For the case of density we show, that $\rho$ is observed to decrease
substantially from its closed-packed value at plane inclinations starting 
from tan$\theta/$tan$\theta_r=1.45$ (in our case tan$\theta_2/$tan$\theta_r=1.7$). 
At very high plane inclinations where a very dilute gas phase is observed 
the equation of state is certainly much different than for the dense flows 
and the above criterion is not relevant.
For the case of velocity $u$, our characterization of $u$ as a function of downstream 
distance demonstrates that there is a healing length of order the size of our inclined 
plane for moderate plane inclinations that complicates a comparison
with the predictions for an acceleration threshold. We conclude that because of 
the limited plane length our data are not complete enough to make a quantitative
prediction of the angle up to which stationary flows exist.

As demonstrated
in other experiments, air can sometimes have a profound effect on the observed 
behavior of the granular flow as in, for example, segregation of vertically-vibrated
granular materials \cite{moch2004,moch2005,zeho2006}, discharging hourglasses 
\cite{wuma1993,muhu2004}, or impact studies \cite{lobe2004}.  
Thus, in order to characterize the properties of granular flow at high inclination 
angles, it is important to understand how air interacts with the granular flow and
to determine the level of fluidization and grain velocity when the role of air drag 
is no longer negligible.  

In this paper, we characterize the flow of relatively mono-dispersed sand 
particles on a rough inclined plane. The experimental apparatus, that allows
for measurements of flow conditions as a function of air pressure, and a 
characterization of the average terminal velocity of individual grains with 
different mean sizes and material composition are presented in Sec.\ \ref{sec:exp}.  
In Sec.\ \ref{sec:results}, we describe our measurements of granular layer 
velocity, height and density as a function of hopper opening $H$. We summarize our 
findings in a phase diagram that includes the transition to a gaseous phase. 
Finally, we conclude with some discussion of the 
implications and further opportunities for the system of granular flow on
an inclined plane.

\section{Experiment}
\label{sec:exp}

We start by characterizing the granular material and describing the experimental 
apparatus used to make quantitative measurements of the flow.  For determining the 
flow properties of the granular layer, we used sand that was sifted with 
300 and 500 $\mu$m sieves to yield a mean diameter of $d=400\ \mu$m. 
We designate this distribution as having a mean of $d=400\ \mu$m and a standard 
deviation of $50\ \mu$m.  We also used finer sand, salt and glass beads to help 
calibrate the air drag effects.

\begin{figure}[ht]
\resizebox{85mm}{!}{
\includegraphics{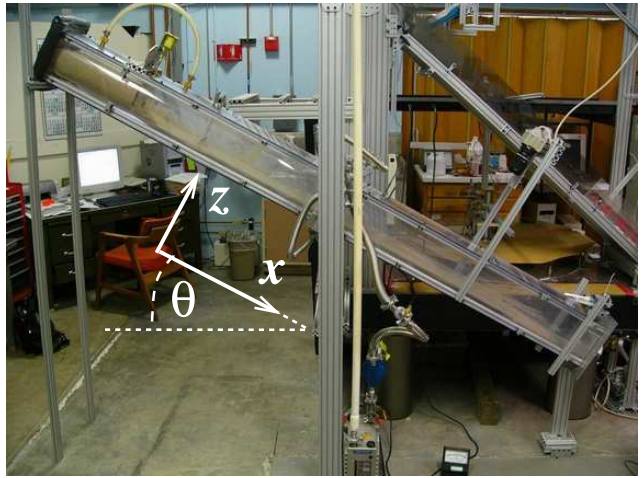}
}
\caption{Color online. Photograph of the experimental setup.
}
\label{setup}
\end{figure}

The experimental setup used for measurements of inclined plane flows
is shown in Fig.\ \ref{setup}. A glass plate with dimensions 
230 cm x 15 cm was set inside a 274 cm long, 20 cm diameter cast-acrylic tube.
The leftmost 40 cm of the tube (full with sand in the image) serves as the hopper.
The surface of the remaining part (190 cm) of the glass plate was made rough 
by gluing one layer of grains onto it or by covering the plate with sandpaper 
that had a characteristic roughness of 190 $\mu$m (80 grit).
The same flow regimes were observed for both surfaces.
Because the surface of the sandpaper was slightly smoother, we observed a small 
(about 10$\%$) increase of the flow velocity compared to the case of
grains glued onto the plate.
The tube was rotatable about the middle so that 
we could set an arbitrary inclination angle $\theta$. 
The whole system could be pumped down to $P \approx$ 0.5 mbar.

The flow velocity $u$ at the surface was determined by analyzing high speed 
(8000 frames per second) recordings. Space time plots were created by taking one 
line of the recordings parallel to the main flow. The Fourier transform of such 
an image (consisting of streaks as traces of particles in the flow) yields the 
average velocity of  surface particles. The thickness of the flow $h$ was 
monitored by the translation of a laser spot formed by the intersection with 
the surface of a laser beam aligned at an angle of $20^o$ with respect
to the inclined plane.

\begin{figure}[ht]
\resizebox{80mm}{!}{
\includegraphics{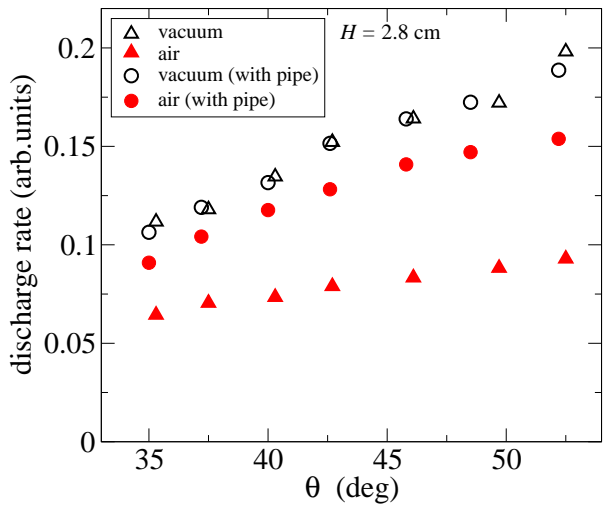}
}
\caption{
Color online. Hopper discharge rate with $H=2.8$ cm vs. $\theta$
at ambient pressure (filled symbols) and at $P$ = 0.5 mbar (open symbols).
The two datasets correspond to the case of  two separated chambers (triangles)
and reduced counterflow by a connecting pipe (circles). 
}
\label{discharge}
\end{figure}

The flow properties were characterized by varying two control parameters: the plane
inclination $34.1^o < \theta < 52.2^o$ and the hopper opening $0.4 < H < 4.4$ cm
({\it i.e.} $10 < H/d < 110$). 
Setting a constant $H$, we observed a slight increase of the hopper discharge 
rate when increasing $\theta$, {\it e.g.}, see Fig.\ \ref{discharge}. The hopper 
discharge rate also depended on the presence of air in the system 
\cite{wuma1993,vedi1997,muhu2004}.
In a typical hourglass geometry, when the hopper and the main chamber are 
separated, the flow of sand faces a counterflow of air as the hopper discharges 
and a pressure difference builds up. Without a connection between the two chambers
in our experiment, the hopper discharge rate was significantly smaller
(by about $50\%$) in the presence of air when compared to the case in vacuum.
By connecting the two chambers with a flexible tube, the counterflow was 
substantially reduced, and the discharge rate was about $80-90\%$ of the discharge 
rate observed in vacuum (Fig.\ \ref{discharge}). 
All the measurements presented here were done for the case of reduced counterflow.

To set the stage for our discussions of air effects in inclined plane flows, we 
introduce some results related to the interaction of granular particles with air.
One way in which air can affect a granular flow is through the drag force of air 
acting on an isolated particle. Given a spherical particle of radius $R$ 
(diameter $d$) and density $\rho_p$ falling under gravity $g$ in air with dynamic 
viscosity $\mu$ and density $\rho_F$, the equation of motion is 
\begin{equation}
{dv \over dt} + {9 \mu v c(v) \over 2 \rho_p R^2} = g
\label{eqn:EOM}
\end{equation}
where the buoyancy term is ignored since $\rho_p >> \rho_F$ and $c(v)$ is a 
turbulent drag correction \cite{clgr1978} $c(v) = 1 + 0.15(vd/\nu)^{2/3})$ with 
$\nu = \mu/\rho_F$. The turbulent drag coefficient $c(v)$ yields a good 
approximation for the range of velocities that are of interest here. 
The terminal velocity $v_T$ is obtained by setting $dv/dt=0$ and solving numerically 
the equation
\begin{equation}
 0.15({2R \over \nu})^{2/3}v^{5/3} + v - {2 \over 9} {\rho_pR^2g \over \mu} = 0
\label{eqn:EOM2}
\end{equation}

The air drag on the individual grains is characterized by the Reynolds number 
$Re \equiv vd/\nu$ of a particle of diameter $d$ falling at velocity $v$.  
Taking spherical particles with $R=0.2$ mm and with the density of sand 
$\rho_p=2.5$ g/cm$^3$, by solving Eq. (\ref{eqn:EOM2})
numerically we obtain  $v_T=3.3$ m/s with a corresponding $Re = 90$.

\begin{figure}[ht]
\resizebox{85mm}{!}{
\includegraphics{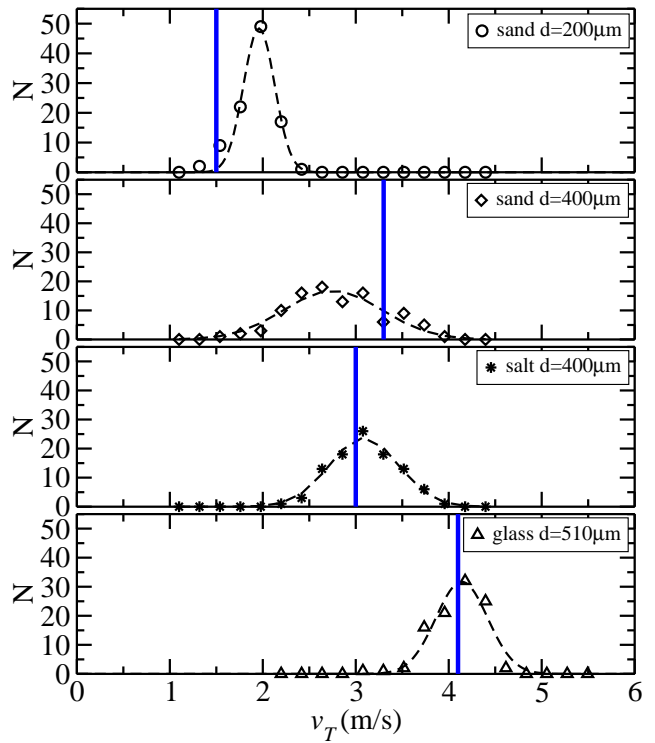}
}
\caption{
Color online. Distribution of experimental values of $v_T$ for different granular materials:
fine sand ($d=200\pm50\ \mu$m),  sand ($d=400\pm50\ \mu$m), salt ($d=400\pm50\ \mu$m), and
glass beads ($d=510\pm50\ \mu$m). The dashed lines are Gaussian
fits to the data, and the solid vertical lines are the calculated values of $v_T$.
}
\label{terminalvelo}
\end{figure}

The values of $v_T$ were also measured by dropping particles from 2.7 m. 
All of the particles reached a terminal velocity where the change in their 
velocities was below $1\%$ over a 7 cm interrogation window. The velocity 
distribution reflecting the variations in particle size and shape is presented for 
100 particles in Fig.\ \ref{terminalvelo}. The average value of the terminal velocity
for the 400 $\mu$m sand is $v_T=2.7$ m/s. 
The measurements were also done for fine sand $d=200\pm50 \ \mu$m, salt 
$d=400\pm50\ \mu$m and glass beads $d=510\pm50\ \mu$m, yielding the 
the velocity distributions presented in Fig.\ \ref{terminalvelo}.
The corresponding mean values of $v_T$ are $1.8$ m/s, $3.0$ m/s and $4.0$ m/s, 
respectively. The calculated values of $v_T$ are indicated with a vertical line 
for each case. The calculated values of $v_T$ are close to the measured 
values with deviations between data and theory presumably resulting from non-spherical 
shapes and/or uncertainties in the mean particle diameter.
For example, the deviations observed for the two sets of sand can be
consistently explained with non-centered size distributions.

\begin{figure}[ht]
\vspace*{0.5cm}
\resizebox{80mm}{!}{
\includegraphics{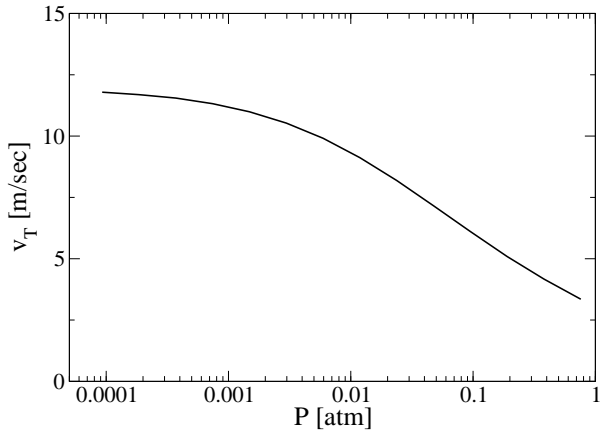}
}
\caption{
Calculated terminal velocity of spherical particles  with $d=400\ \mu$m and 
$\rho = 2.5$ g/cm$^3$ vs. $P$.  
}
\label{vt-pressure}
\end{figure}

Having determined that our calculation of $v_T$ agrees well with the 
measured data, we used Eq.\ \ref{eqn:EOM2} to estimate the air drag reduction
and the resulting increase in $v_T$  as a function of air pressure.  
The dynamic viscosity $\mu$ of air is almost constant as a function of air 
pressure $P$.  Using an ideal gas relationship between $\rho_F$ and P and 
Eq.\ \ref{eqn:EOM2}, one obtains $v_T$ (and $Re$) as a function of $P$.
As seen in Fig.\ \ref{vt-pressure}, $v_T$ is expected to increase by a factor 
of 4 when $P$ decreases to 0.5 mbar. If air drag plays a significant role in 
determining the granular flow state, the substantial decrease in air drag 
under vacuum will reveal that effect.  Here we ignore the expected
further increase in the terminal velocity at very low pressures when the
air mean free path is comparable to the particle diameter.

Another approach to estimating the effect of air drag on a flowing granular 
layer is to assume that the granular layer is similar to a fluid and 
that the air forms a boundary layer between air at large distance with zero 
velocity and the granular layer flowing with a ``surface velocity'' $v$.  
A Prandtl boundary layer description yields a boundary layer thickness 
$\delta = 3.4 (\nu x/v)^{1/2}$ where $x$ is the downstream distance.  
The force per unit  area $f_A$ exerted on the flow by the entrained air is 
approximately $0.66\ (\rho_F \mu v^3/x)^{1/2}$. The relevant parameters of 
the air are $\mu = 1.789 \ \cdot 10^{-5}$ Pa s, $\rho_F = 0.96$ kg/m$^3$ 
(for local pressure $P \approx 0.76 P_{ATM}$), and $\nu = 0.19$.  
Taking a distance $x = 100$ cm and $v= 300$ cm/sec, one obtains 
$\delta \approx 0.2$ cm and $f_A \approx 0.01$ N/m$^2$ which is
less than 0.1$\%$ of the gravitational force $-$ a negligible effect.
The surface of a granular flow, however, is not as sharply defined as it is for 
liquid flows. If there were substantial fluidization of the granular layer,
the cumulative drag force experienced by the low density particles might play a 
role in the instabilities of the homogeneous flow.

\section{Results and Discussion}
\label{sec:results}

\subsection{Phase diagram}
\label{subsec:phasediag}

\begin{figure}[ht]
\resizebox{85mm}{!}{
\includegraphics{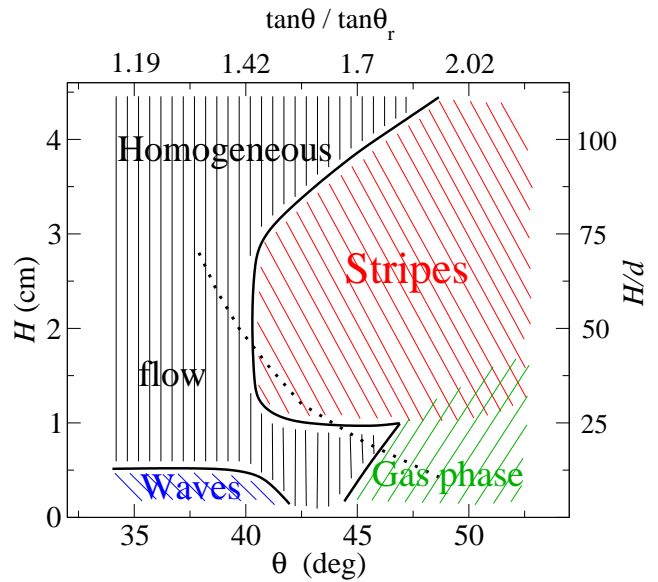}
}
\caption{Color online. Phase diagram of the flow regimes in the phase space defined by 
$H$ and $\theta$. The dotted line divides the diagram into
two regions: low flow rates and not too steep planes where  $x-$invariant 
steady flow is observed and
higher flow rates and faster flows where the flow is still accelerating as measured 
at $x=1$ m. 
The ratios $\tan\theta/\tan\theta_r$ and $H/d$ are labeled on the top and right axes, 
respectively. 
}
\label{phasediagram}
\end{figure}

Based on qualitative observations of the flow and on quantitative measurements 
presented below, we can separate the granular flow into regions with certain 
characteristic features.  Such a phase diagram identifying the different flow 
regimes as a function of the two control parameters $H$ and $\theta$ is shown in 
Fig.\ \ref{phasediagram}. The ratios tan$\theta$/tan$\theta_r$ and $H/d$ are 
also shown on the top and right axes, respectively, where $\theta_r = 30.5$ is 
the independently measured bulk angle of repose for the granular material 
used here.  The boundaries are neither precisely defined nor indicative of sharp
transitions between different flow regimes. Further, this phase diagram does not 
capture the convective nature of the granular flow in that features or flow 
properties may evolve over length scales comparable to the length of the inclined
plane. Nevertheless, identification of the general regimes are useful in setting
the stage for more complex issues raised below.

At very slow flow rates and low plane inclinations the homogeneous flow is unstable
and organizes itself into waves. The properties of such waves have been studied 
extensively \cite{fopo2003}. By increasing the flow rate and staying in the 
range of not very steep plane inclinations (up to about $40^o$), the flow becomes 
homogeneous. For steeper planes an instability occurs leading to a pattern 
consisting of lateral stripes \cite{fopo2001,fopo2002}. A detailed characterization 
of the stripe state in our system will be presented elsewhere \cite{boec2005}.
The dashed line divides the diagram into two regions: accelerating and 
$x-$invariant steady flows. Above the dashed line the acceleration of 
the flow (averaged over the range of $60<x<140$ cm) was more than 5\% of  g sin$ \theta$. 
This boundary depends on the distance downstream $x$ at which the 
acceleration was measured as discussed in more detail in Sec. \ref{subsec:velocity}.
The phase diagram was found to be qualitatively the same for 
flow at ambient pressure and flow at low pressure with $P=0.5$ mbar.

\subsection{Flow thickness and density}
\label{subsec:thickness}

In this section, we describe our procedure for measuring layer height and
mean density of the flowing layer.  
The flow thickness was measured by laser deflection.
\begin{figure}[h]
\vspace*{0.1cm}
\resizebox{60mm}{!}{
\includegraphics{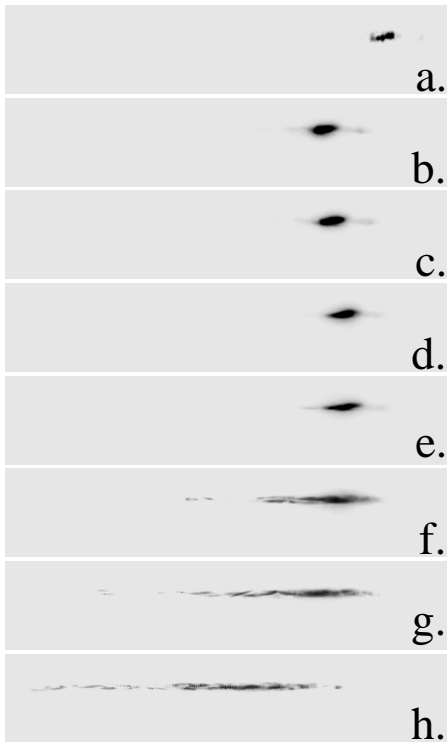}
}
\caption{Image of the laser spot for (a) no flow and
for flow at inclination angles $\theta$ of
(b) $35.0^o$,
(c) $37.2^o$,
(d) $40.0^o$,
(e) $42.6^o$,
(f) $45.8^o$,
(g) $48.5^o$ and
(h) $52.2^o$.
$H=0.4$ cm and  $x=79$ cm.
}
\label{laserspot}
\end{figure}
The plane of the images shown in Fig.\ \ref{laserspot} was parallel to the inclined 
plane. A laser beam projected from the left hand side of the image at an angle of 
$20^o$ with respect to the plane of the image produced a localized laser spot. 
The first image, Fig.\ \ref{laserspot}a, shows the case of the empty chamber
without flow, where the laser beam was reflected by the rough surface of 
the inclined plane. The other images (b-h) were taken in the presence of flow 
at a constant hopper opening $H=0.4$ cm and downstream distance $x=79$ cm
with increasing  $\theta$ as we go 
from (b-h). The laser beam was reflected from the particles at the surface of 
the flowing layer. The horizontal shift of the laser spot with respect to 
image (a) measured the thickness of the flowing layer. This measurement is 
straightforward when the laser spot did not change its shape, {\it i.e.}, 
the surface of the flow was well defined (Figs.\ \ref{laserspot}b-e).
The first sign of decreasing flow density can be seen in Fig.\ \ref{laserspot}f 
where the shape of the reflected laser spot changed significantly.  
The spot spreading became more pronounced with increasing $\theta$
(Figs.\ \ref{laserspot}g-h).

\begin{figure}[ht]
\resizebox{80mm}{!}{
\includegraphics{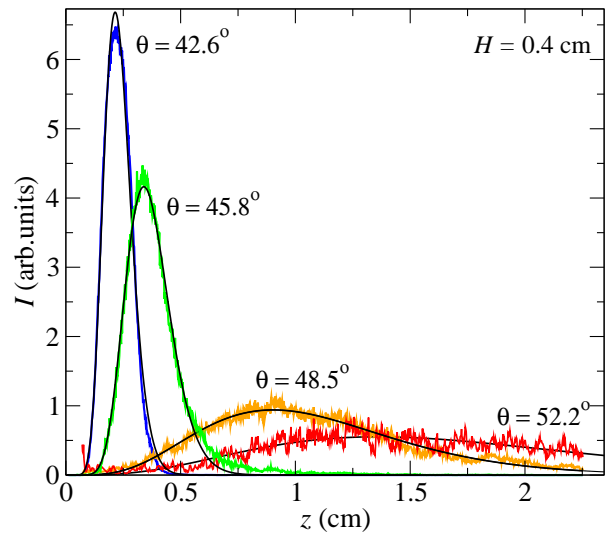}
}
\caption{
Color online. Time-averaged light intensity I(z) vs. $z$ for $\theta$ values indicated.
$H=0.4$ cm,  $x=155$ cm. Solid lines are curve fits of the form $az^be^{-cz}$.
}
\label{heightintensityprofile}
\end{figure}

To get a better measure of the flow thickness in this regime the light intensity 
$I$ detected at a height $z$ above the inclined plane was integrated over 200 
equally-spaced images taken over a time period of 3.3 s. The $I(z)$ curves are 
shown in Fig.\ \ref{heightintensityprofile} corresponding to the last four plane 
inclinations of Fig.\ \ref{laserspot}.  Because of the increasing fluctuations
in the laser spot intensity for the higher $\theta$, averaging over 
many images was necessary. 

For the relatively dense regime (with a compact laser spot), the position of the 
center of mass of the laser spot was taken as the flow thickness $h$.
In the very dilute regime we integrate the $I(z)$ curves and 
define $h$ as the height below which 80$\%$ of the flowing grains were detected.

\begin{figure}[ht]
\resizebox{80mm}{!}{
\includegraphics{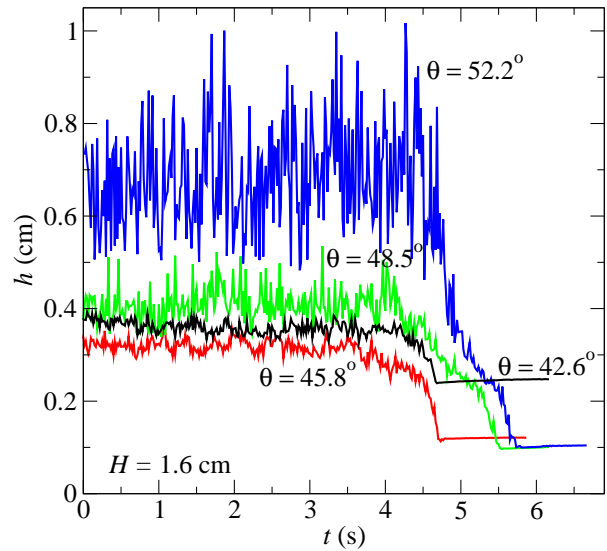}
}
\caption{
Color online. Flow thickness $h$ vs. $t$ for plane inclinations
$\theta=42.6^o$, $45.8^o$,  $48.5^o$ and $52.2^o$.
$H=1.6$ cm, $x=155$ cm. Flow was stopped  at $t \approx 4$ s.
}
\label{height-time}
\end{figure}

The flow density was measured in the following manner.
A stationary flow was established  and was maintained for about 4 seconds. The 
flow was then ``frozen'' by rapidly decreasing the plane inclination.
\begin{figure*}[ht]
\includegraphics[width=15cm]{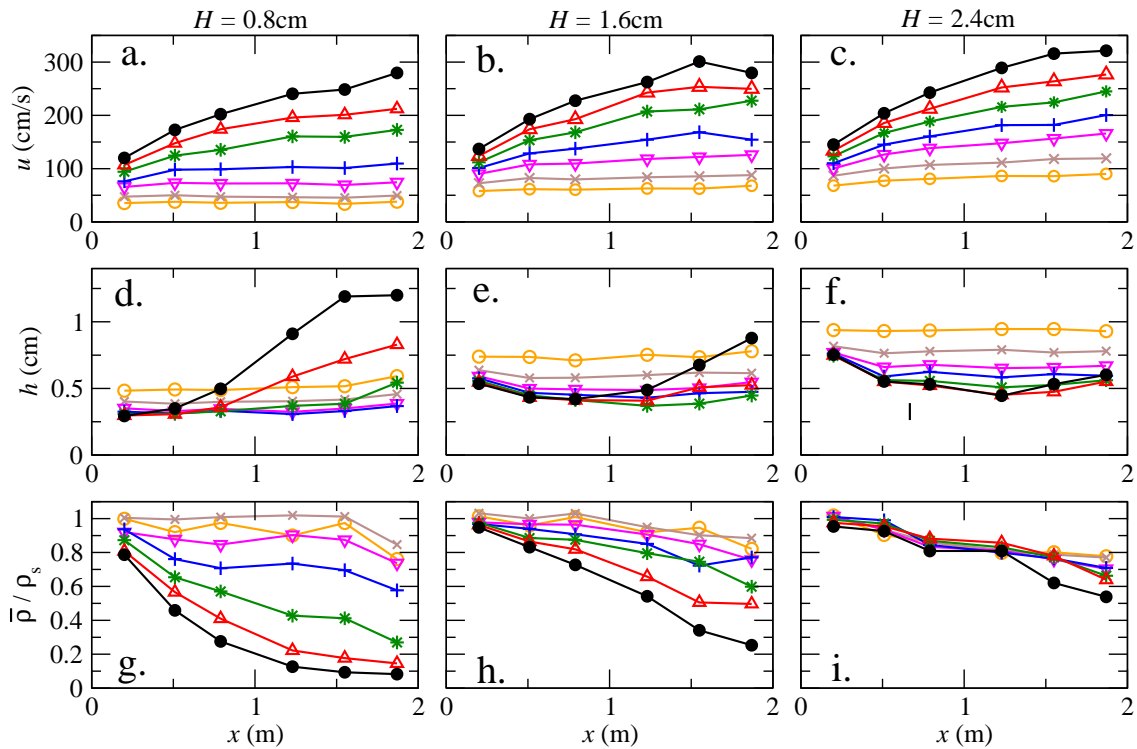}
\caption{
Color online. Velocity $u$ (a-c) thickness $h$ (d-f) and calculated normalized flow density
$\overline{\rho}/\rho_s$ (g-i) as a function of $x$ the distance measured from the hopper
gate for three values of the hopper opening $H=0.8$ cm, $H=1.6$ cm and $H=2.4$ cm.
The symbols correspond to the plane inclinations
$\theta$ of $35.0^o$ ($\circ$), $37.2^o$ (x), $40.0^o$ ($\nabla$),
$42.6^o$ (+), $45.8^o$ ($\star$), $48.5^o$ ($\triangle$)
and $52.2^o$ ($\bullet$).
}
\label{velo-x-theta}
\end{figure*}
The time evolution of the center of mass of the laser spot is shown
in Fig.\ \ref{height-time} for the same set of plane 
inclinations as in Fig.\ \ref{heightintensityprofile}. 
The flow height fluctuated
in the first 4 seconds with larger fluctuations in $h$ for
higher values of $\theta$. As the flow 
stopped, the height dropped to the height corresponding to the static 
(nearly random closed packed) material. The height change between the 
stationary flow and the frozen state reflected the 
mean density change which became larger with increasing $\theta$, see
Fig.\ \ref{height-time}. 
The ratio of the static height and the height of the stationary flow gives
the normalized depth-averaged density of the flow $\overline{\rho}/\rho_s$, 
where $\rho_s$ is the static density. This measurement of height and density
averages over some distance in $x$ because the flow does not stop instantaneously.
Assuming an exponential decay of the velocity and a total stopping time of about
0.5 sec from the data in Fig.\ \ref{heightintensityprofile}, one obtains an averaging
length of between 10 and 50 cm.  Below we describe a more local albeit less
precise measure of density derived from measurements of $u$ and $h$.

\subsection{Flow characterization as a function of $\theta$}
\label{subsec:velocity}

Granular material exits the hopper at low velocity with height $H$ and a density
$\rho_s$ that is close to that of a random close-packed state.  The material 
accelerates, thins and becomes less dense as the interaction of the grains 
with the rough bottom surface partially fluidizes the granular state.
At low inclination angles, the system reaches an an $x-$invariant steady 
state after some healing length $\xi$ where layer height $h$, mean velocity $u$, 
and mean density $\rho$ do not change with downstream distance $x$. For $\theta$ 
larger than about $40^o$, the flow is not stationary as a function of $x$  and more 
complicated states are observed. The flow appears to become stationary on healing
lengths of order the plane length $L$ for large values of $\theta$ as described below.

All measurements of $u$ and $h$ were repeated at six locations at a distance $x$
measured from the hopper in the range of $20$ cm $<x< 187$ cm. The flow velocity
$u$ and layer height $h$ are shown as functions of $x$ for three hopper openings
$H=$  0.8 cm,  1.6 cm and 2.4 cm in Figs.\ \ref{velo-x-theta}a-f.

\begin{figure}[ht]
\resizebox{85mm}{!}{
\includegraphics{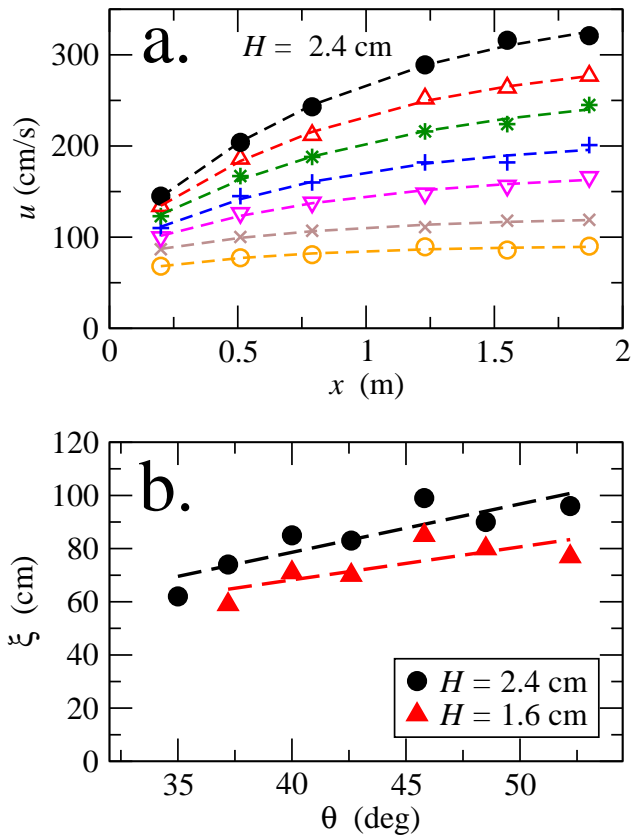}
}
\caption{
Color online. (a) Fits to the $u(x)$ data taken at $H=2.4$ cm.
The symbols correspond to the plane inclinations
$\theta$ of $35.0^o$ ($\circ$), $37.2^o$ (x), $40.0^o$ ($\nabla$),
$42.6^o$ (+), $45.8^o$ ($\star$), $48.5^o$ ($\triangle$)
and $52.2^o$ ($\bullet$).
(b) Healing length $\xi$ as a function of $\theta$ for $H=1.6$ cm and $H=2.4$ cm.
}
\label{healing-length}
\end{figure}

For higher plane inclinations the flow did not reach a stationary state over the plane
length $L$. The evolution of the granular flow can be partially understood by an
analysis of $u$ as a function of downstream distance $x$ for different $\theta$.
The flow is driven by the gravitational force $\rho g \sin \theta$ and damped by
dissipation forces, {\it e.g.}, friction and/or inelastic collisions.
In the simplest approach we can assume that the flow would approach a terminal
velocity $u_f$ like $u(x) = u_f -(u_f-u_0)e^{-x/\xi}$ where $\xi$ corresponds
to the healing length of the flow.
Fitting the data in Fig.\ \ref{velo-x-theta}a-c to this
form allows for the determination of $\xi$ and the velocities $u_0$ and $u_f$ for
different $H$ and $\theta$. In Fig.\ \ref{healing-length}a, we show data for $u$
and fits to the exponential form. The data are well fit by exponential saturation.
The resulting values of $\xi$ are shown in Fig.\ \ref{healing-length}b. as a function
of $\theta$ for $H = 1.6$ cm and $H = 2.4$ cm. The healing lengths are of order
$L/2$ for $\theta > 43^o$.  Because the data show little variation with $x$ for
smaller $\theta$, our fits almost certainly overestimate $\xi$ in that region.

\begin{figure}[ht]
\resizebox{85mm}{!}{
\includegraphics{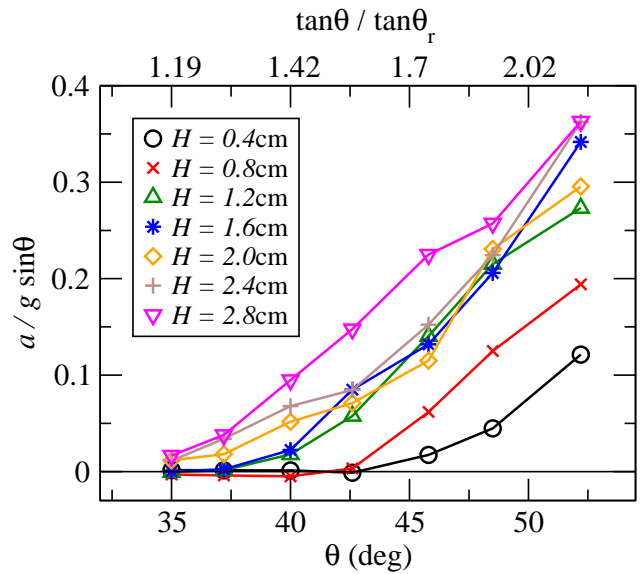}
}
\caption{
Color online. Acceleration averaged over $60<x<140$ cm
for  $H$ values of 0.4 cm ($\circ$), 0.8cm (x), 1.2 cm ($\triangle$),
1.6 cm ($\star$), 2.0 cm ($\diamond$),  2.4 cm (+) and 2.8 cm ($\nabla$).
}
\label{accel-theta}
\end{figure}

From information about the surface velocity $u$ and the layer thickness $h$, additional
information can be inferred about the granular layer despite the inability to 
directly measure the $z$ dependence of density and velocity. If we assume that the 
$u(z)$ and $\rho(z)$ profiles do not change their character along $x$, then by 
conservation of mass one has that the mass flux per unit channel width  
is $F= h \overline{\rho u}$ where the overbar denotes a depth average.
Numerical simulations \cite{sier2001} suggest that for at least 
some range of $\theta$, $\rho$ is almost constant over the depth. With that 
assumption, we have that $F \approx \overline{\rho} h \overline{u}$.
Thus, as a first approximation, plotting $F/u h$ as a function of $x$ provides 
information about the downstream evolution of the mean density $\overline{\rho}$.
The degree to which this is a good approximation depends on the details of
$\rho(z)$ and $u(z)$. The slight difference in $\overline{\rho}$ between the values 
estimated this way and measured by a more precise method presented  
Sec. \ref{subsec:thickness}, can result from a nonuniform ($z$ dependent 
density) at faster flows.
Because we do not know the absolute value for $F$, we normalize the resulting values 
of $\overline{\rho}$ assuming that the flow density for the lowest plane inclination 
was near to the density of the static packing $\rho_s$ near the hopper. 
For the other values of $\theta$ we assumed a $60\%$ increase of the hopper flow rate 
between $\theta=35.0^o - 52.2^o$ for all values of $H$. 
The resulting normalized mean density $\overline{\rho}/\rho_s$ is shown in 
Figs. \ref{velo-x-theta}g-i as a function of $x$. One sees that the 
density drops rapidly as a result of increasing velocity for higher plane inclinations
especially for the case of lower incoming flow rates. A more precise characterization
of the density change using the method described in Section \ref{subsec:thickness}, 
made at one downstream location $x=155$ cm, will be presented below. Results of the 
two methods agree within the uncertainties inherent in each approach.

Using the $u(x)$ curves the acceleration of the flow was obtained by assuming that
$a=du/dt \approx u du/dx$.  The average value of $a$ over the range 
60 cm $<  x <$ 140 cm is presented as a function of $\theta$ in 
Fig. \ref{accel-theta}.  
At higher plane inclinations, there is a significant acceleration for each 
value of $H$ whereas for the lowest value of $\theta$ the flow is 
$x-$invariant for all the $H$ values we investigated.
The value of $H$ corresponding to $a=0.05 g \sin \theta$ is indicated as a function 
of $\theta$ in Fig. \ref{phasediagram} with a dotted line as a boundary between 
the accelerating and $x-$invariant steady regimes measured at this distance 
from the hopper.

In the following we analyze the velocity $u$, flow thickness $h$ and dimensionless 
mean flow density $\overline{\rho}/\rho_s$ as a function of $\theta$ and $H$ at the
location $x=155$ cm below the hopper. Parts of our data correspond to accelerating
non-stationary flow. Since we are reporting averages obtained near the channel 
center, we need to determine the degree to which those averages depend on the 
lateral homogeneity of the flow. Thus, we first indicate some measure
of the lateral flow structure. 

\begin{figure}[ht]
\resizebox{85mm}{!}{
\includegraphics{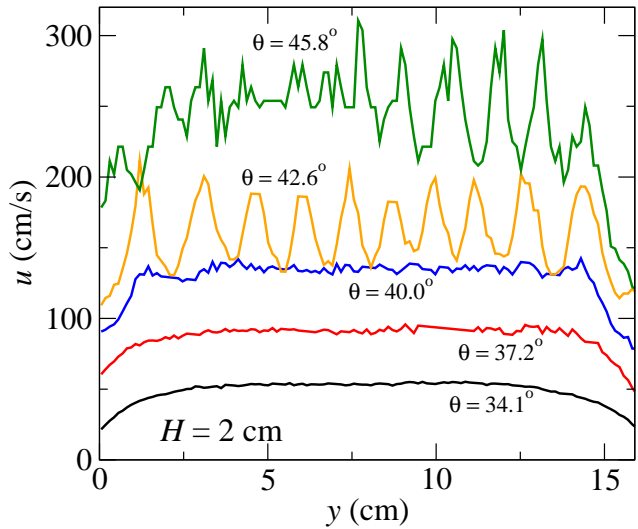}
}
\caption{
Color online. Surface velocity profiles $u(y)$ of the flow  for plane inclinations of
$\theta=34.1^o$, $37.2^o$, $40.0^o$, $42.6^o$ 
and $45.8^o$.
The data were taken the location $x = 155$ cm below the hopper at $H=2.0$ cm.
}
\label{velo-profiles}
\end{figure}

Lateral velocity profiles $u(y)$ are presented in Fig.\ \ref{velo-profiles}. 
In the homogeneous dense flow regime, the velocity is relatively
constant over the center 12 cm of the channel corresponding to about 75\% of the channel
width.  At the edges of the channel there are boundary layers that arise from 
the friction with the sidewalls. As the sidewalls of the channel have a smooth
surface, friction is significantly less important there when compared to the rough 
plane. The transition to the stripe state (see curves in Fig.\ \ref{velo-profiles} 
with $\theta$ values of 42.6$^o$ and 45.8$^o$) happens at around $\theta = 41^o$. 

\begin{figure}[ht]
\vspace*{0.1cm}
\resizebox{85mm}{!}{
\includegraphics{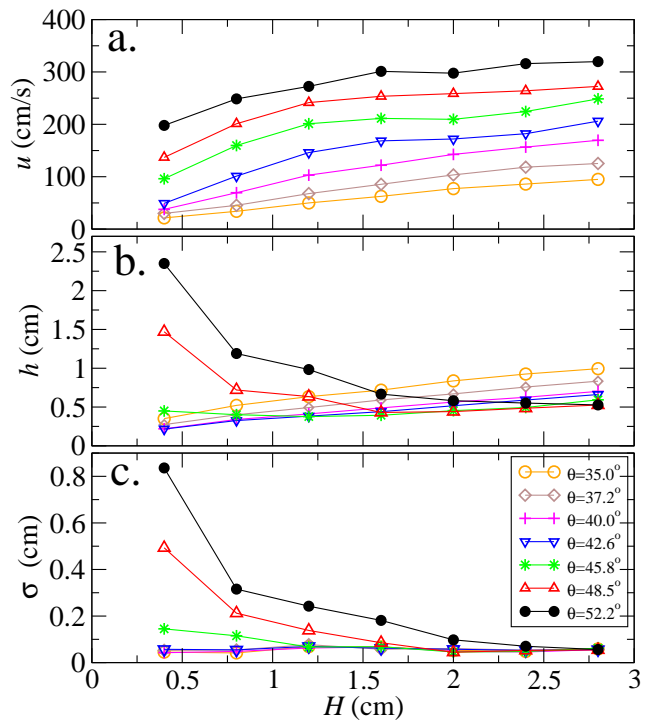}
}
\caption{Color online. The flow velocity $u$, layer thickness $h$ and its standard deviation 
$\sigma_h$ vs. $H$ at $x = 155$ cm.
}
\label{height-velo-spread-opening}
\end{figure}

The surface flow velocity $u$, the height of the flowing layer $h$ and 
its standard deviation defined as 
$\sigma = \sqrt{\frac{\int I(z) (z - h)^2dz}{\int I(z)dz}}$
are shown as a function of $H$ in Fig.\ \ref{height-velo-spread-opening}
as measured at $x= 155$ cm. 

The flow velocity is a monotonically 
increasing function of $H$ for all plane inclinations. In the fluid-like regime 
$h$ is a monotonically increasing function of $H$ as well, whereas $\sigma$ has a 
constant low value reflecting the constant shape of the laser spot. The sign of the
transition into the gaseous phase is the rapid increase of $h$ and $\sigma$
with decreasing $H$ at higher plane inclinations.

\begin{figure}[ht]
\vspace*{0.1cm}
\resizebox{85mm}{!}{
\includegraphics{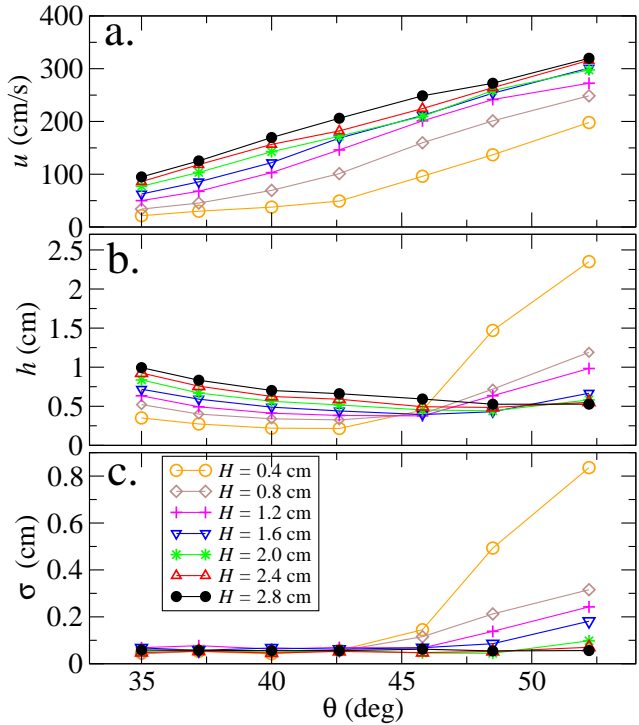}
}
\caption{
Color online. The flow velocity $u$, flow thickness $h$ and its standard deviation $\sigma_h$ 
as a function of  $\theta$ at $x = 155$ cm.
}
\label{height-velo-spread-inclination}
\end{figure}

The same set of data for $u$, $h$ and $\sigma$ is presented in 
Fig.\ \ref{height-velo-spread-inclination} as a function of $\theta$.
Again $u$ is a monotonically increasing function of $\theta$ for each value 
of $H$. In the fluid-like regime (larger values of $H$) the slope of the 
$u(\theta)$ curves is nearly constant. 
At smaller $H$ the slope increases considerably when entering the gaseous regime. 
The flow thickness $h$ decreases with increasing $\theta$ in the fluid-like 
regime meaning that larger flow velocity results in smaller flow thickness.
This effect is actually stronger than Fig.\ \ref{height-velo-spread-inclination}b
indicates, as the incoming flow rate slightly increases with increasing $\theta$ 
(see Fig.\ \ref{discharge}).
Thus, thinning of the flow as a result of larger flow velocities is slightly 
counterbalanced by the growing value of the flow rate by increasing $\theta$.

\begin{figure}[ht]
\resizebox{85mm}{!}{
\includegraphics{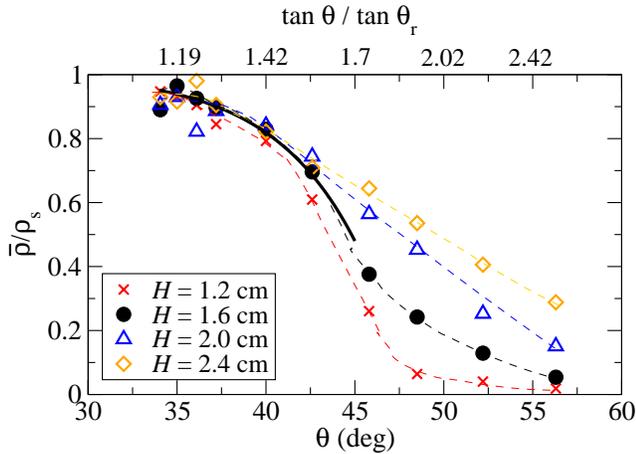}
}
\caption{
Color online. Normalized mean density $\overline{\rho}/\rho_s$ as a function of 
$\theta$ for $H=1.2$ cm (x), $1.6$ cm ($\bullet$), $2.0$ cm ($\triangle$)
and $2.4$ cm ($\diamond$) at  $x = 155$ cm. The dashed lines are guides to the eye,
while the continous line corresponds to $\rho/\rho_s=1-0.52\cdot$tan$^6\theta$. 
}
\label{inclination-density}
\end{figure}

The different flow regimes can best be characterized by the flow density.
As we see in Fig.\ \ref{inclination-density} the normalized mean density 
$\overline{\rho}/\rho_s$ drops continuously with increasing $\theta$. 
The density drop is more dramatic for smaller incoming flow rate than
for thicker flows.

The {\it first regime} corresponds to dense flows (slightly above the angle of repose)
and is often characterized by the Pouliquen flow rule, where the depth averaged 
velocity $\overline{u}$ is proportional to $h^{3/2}$ \cite{po1999,fopo2003}.  
This regime is reported to exist for the plane inclinations 
$\tan\theta/\tan\theta_r < 1.45$, and the dense flow can be unstable with respect to 
the formation of waves \cite{fopo2003}. According to our measurements the mean density
in this regime is slightly decreasing with $\theta$ but always stays larger than 
$\overline{\rho}/\rho_s = 0.8$. 
According to a recent theory by Jenkins \cite{je2006} the density should 
decrease as $\rho/\rho_s=1-B\cdot$tan$^6\theta$. 
A reasonable fit is obtained by this formula for a range of $H$ up to the 
plane inclination $\tan\theta/\tan\theta_r < 1.5$ yielding $B=0.52$.
In the {\it second regime} falling in the range of 
$1.45< \tan\theta/\tan\theta_r < 2.4$, a stripe pattern can be observed with an average
density of $0.2< \overline{\rho}/\rho_s < 0.8$. The detailed characterization of the 
stripe structure is beyond the scope of this paper.
In the {\it third regime}, the stripe structure disappears as the flow gets very 
fluidized with an average density of $\overline{\rho}/\rho_s<0.2$. 
A dramatic density decrease was observed for lower incoming flow rates, 
typically at $\tan\theta/\tan\theta_r > 1.6$, yielding a gas-like phase where the
density was less than $5\%$ of $\rho_s$. We characterize this transition 
in more detail below.

\begin{figure}[ht]
\vspace*{0.1cm}
\resizebox{85mm}{!}{
\includegraphics{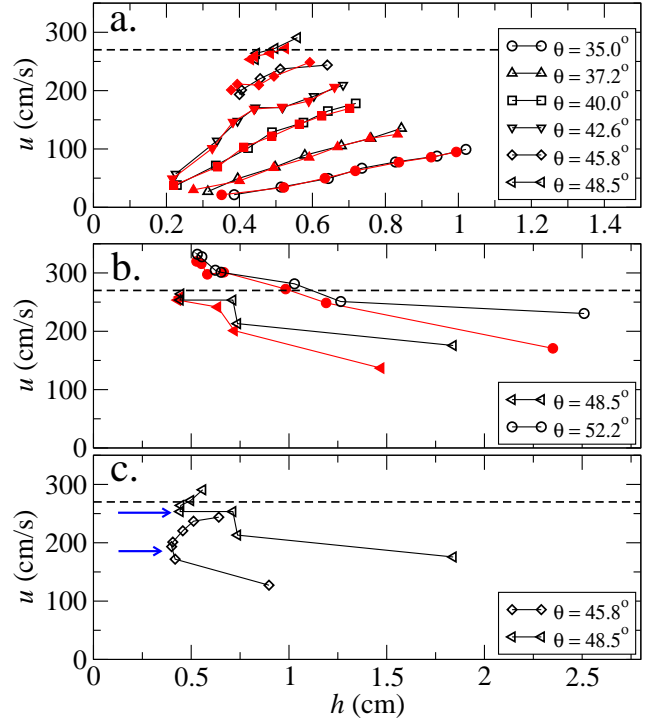}
}
\caption{Color online. The flow velocity $u$ vs. $h$ for a set of plane inclinations in the 
presence of air (filled symbols) and in vacuum (open symbols) in the
(a) fluid-like phase, (b) gaseous phase.
The transition between the above regimes is indicated in (c) by horizontal arrows.
Data taken at $x = 155$ cm.
The horizontal dashed lines correspond to $v_T$.
}
\label{height-velo}
\end{figure}

The surface flow velocity $u$ as a function of height $h$ is presented
in Fig.\ \ref{height-velo}a for the fluid-like regime. 
Note that only the first curve (at $\theta=35.0^o$) corresponds to stationary 
flows, the other curves (partly) fall already in the range of accelerating flows.
Two sets of data 
were taken: (i) in the presence of air (filled symbols) and (ii) at low 
pressure $P=0.5$ mbar (open symbols). In this regime the two sets of data match 
implying that air drag does not become important even at the 
fastest flows where the grain velocity is close to the value of the terminal 
velocity measured in free fall. 
In the gaseous regime, however, (see Fig.\ \ref{height-velo}b)
particle velocities measured at ambient pressure are slightly smaller
than at low pressure. Thus, in the gas phase the average 
interparticle distance is considerably increased  and the contribution of air drag 
to the dissipation becomes important.

The changing nature of the flow is visualized in Fig.\ \ref{height-velo}c 
where all the data points are presented for two plane inclinations for the 
two regimes and the transition between them.
The sharply different tendency of the $u(h)$ curves in Fig.\ \ref{height-velo}a 
and Fig.\ \ref{height-velo}b reflects the nature of the transition between the 
fluid-like and gaseous phases. 
For $\theta=48.5^o$ the larger flow rates give
rise to dense flows (Fig.\ \ref{height-velo}a) where increasing $H$
 leads to increasing $h$ and faster flow.
At lower values of the hopper flow rate decreasing $H$ leads to increasing levels of 
fluidization and to a transition to the gaseous phase. 
In this regime, see Fig.\ \ref{height-velo}b, the density of the flow rapidly 
decreases with decreasing $H$ and the measured thickness rapidly increases.
Contrary to trends for the fluid-like phase, the leftmost data 
points of the curves in Fig.\ \ref{height-velo}b correspond to the highest 
hopper flow rates and the rightmost ones to the lowest hopper flow rate.
Thus, the flow is not simply determined by ($\theta$, $h$, $x$) as is 
the case for dense flows. For a given $h$, $\theta$ and $x$, two solutions exists, 
one corresponding to a dense phase and the other to a very dilute one, depending on 
the flow rate.

A recent linear stability analysis indicated, that the air drag could play a 
role in the development of stripe patterns in the fluid-like regime 
\cite{arts2006,ar2004}.
According to our findings the density of the flow for the stripe state was in 
the range of $0.2< \overline{\rho}/\rho_s < 0.8$, corresponding to the regime 
where the role 
of the air drag is minor. The  flow properties are not affected even at 
flow velocities near the terminal velocity in free fall. At the transition 
to the gaseous state the stripes disappear and the rapid decrease of the 
flow density leads to visible effects of air drag.

In summary, we have presented a detailed description of granular flow on a rough 
inclined plane - one of the most commonly used model systems for granular dynamics -
concentrating on the fast flow regime. We developed a method to measure the depth
averaged normalized flow density $\overline{\rho}/\rho_s$ and characterized the 
flow regimes as a function of $\overline{\rho}/\rho_s$. We have characterized the 
transition to a very dilute gaseous phase that takes place by increasing plane 
inclination and decreasing incoming flow rate. 
By measuring the flow properties at ambient pressure and at low pressure
($P\approx0.5$ mbar) in a vacuum flow channel we have shown that the dissipation 
by the air drag (with respect to the other dissipational processes) is 
non-negligible only in the dilute gas-like phase.
Other implications of this work involves the possibility of linking changes in the flow
structure (such as the stripe state) as a function of plane inclination to the change 
in the depth averaged normalized flow density $\overline{\rho}/\rho_s$.
 
This work was funded by the US Department of Energy (W-7405-ENG).
The authors benefited from discussions with I.S. Aranson.
T.B. acknowledges support by the Bolyai J\'anos Scholarship of the Hungarian
Academy of Sciences and the Hungarian Scientific Research Fund (Contract No. 
OTKA-F-060157).


\begin{thebibliography}{99}

\bibitem{an2001}
C. Ancey, Phys. Rev. E {\bf 65}, 011304  (2001).

\bibitem{ra2003}
J. Rajchenbach, Phys. Rev. Lett. {\bf 90}, 144302 (2003).

\bibitem{bede2003}
G. Berton, R. Delannay, P. Richard, N. Taberlet and A. Valance, Phys. Rev. E {\bf 68}, 051303 (2003).

\bibitem{ra2005}
J. Rajchenbach, J. Phys.: Condens. Matter {\bf 17}, S2731 (2005).

\bibitem{ra2004}
J. Rajchenbach, Eur. Phys. J. E {\bf 14}, 367 (2004).

\bibitem{gdrmidi2004}
GDR MiDi, Eur. Phys. J. E {\bf 14}, 341 (2004).

\bibitem{humo1984}
O. Hungr and N.R. Morgenstern, G\'eotechnique {\bf 34}, 405 (1984).

\bibitem{fopo2001}
Y. Forterre and O. Pouliquen, Phys. Rev. Lett. {\bf 86}, 5886 (2001).

\bibitem{po1999}
O. Pouliquen,  Phys. of Fluids {\bf 11}, No.3, 542 (1999).

\bibitem{fopo2003}  Y. Forterre and O. Pouliquen, J. Fluid Mech. {\bf 486}, 21 (2003).

\bibitem{sa1979}  S.B. Savage, J. Fluid Mech. {\bf 92}, 53 (1979).

\bibitem{hawa2000}
D.M. Hanes and O.R. Walton, Powder Technology {\bf 109}, 133 (2000).

\bibitem{azch1999}
E. Azanza, F. Chevoir and P. Moucheront, J. Fluid. Mech. {\bf 400}, 199 (1999).

\bibitem{go2003}
I. Goldhirsch, Ann. Rev. Fluid. Mech. {\bf 35}, 267 (2003).

\bibitem{jana1996} H.M. Jaeger, S.R. Nagel  and R.P. Behringer,
Rev. Mod. Phys. {\bf 68} 1259 (1996).

\bibitem{arts2006}
I.S. Aranson and L.S. Tsimring,
Rev. Mod. Phys. {\bf 78} 641 (2006).†

\bibitem{si2005}
L.E. Silbert, Phys. Rev. Lett. {\bf 94}, 098002 (2005).

\bibitem{da2000}
A. Daerr,  Ph.D. thesis, University of Paris VII (France), (2000).

\bibitem{fopo2002}
Y. Forterre and O. Pouliquen, J. Fluid Mech. {\bf 467}, 361 (2002).

\bibitem{boec2005}
T. B\"orzs\"onyi and R.E. Ecke, Unpublished.

\bibitem{sier2001}
L.E. Silbert, D. Ertas, G.S. Grest, T.C. Halsey, D. Levine and S.J. Plimpton,
Phys. Rev. E {\bf 64}, 051302 (2001).

\bibitem{sila2003} L.E. Silbert, J.W. Landry and G.S. Grest,
Phys. of Fluids {\bf 15}, 1 (2003).

\bibitem{jofo2005}
P. Jop, Y. Forterre and O. Pouliquen, J. Fluid Mech., {\bf 541}, 167 (2005).

\bibitem{jofo2006}
P. Jop, Y. Forterre and O. Pouliquen, Nature, {\bf 441}, 727 (2006).

\bibitem{moch2004}
M.E. M\"obius, X. Cheng, G.S. Karczmar, S.R. Nagel and H.M. Jaeger,
Phys. Rev. Lett. {\bf 93}, 198001 (2004).

\bibitem{moch2005}
M.E. M\"obius, X. Cheng, P. Eshuis, G.S. Karczmar, S.R. Nagel and H.M. Jaeger,
Phys. Rev. E {\bf 72}, 011304 (2005).

\bibitem{zeho2006}
C. Zeilstra, M.A. van der Hoef, and J.A.M. Kuipers,
Phys. Rev. E {\bf 74}, 010302(R) (2006).

\bibitem{wuma1993}
X-I. Wu, K.J. Maloy, A. Hansen, M. Ammi, and D. Bideau, Phys. Rev. Lett. {\bf 71}, 1363 (1993).

\bibitem{muhu2004}
B.K. Muite, M.L. Hunt, and G.G. Joseph Phys. of Fluids {\bf 16}, 3415  (2004).

\bibitem{lobe2004}
D. Lohse, R. Bergmann, R. Mikkelsen, C. Zeilstra, D. van der Meer, M. Versluis,
K. van der Weele, M. van der Hoef and H. Kuipers
Phys. Rev. Lett. {\bf 93}, 198003 (2004).

\bibitem{vedi1997}
C. T. Veje and P. Dimon, Phys. Rev. E {\bf 56}, 4376 (1997).

\bibitem{clgr1978}
R. Clift, J.R. Grace, and M.E. Weber, Bubbles, Drops, and Particles (Academic
Press, New York, 1978).

\bibitem{je2006}
J.T. Jenkins, Phys. of Fluids {\bf 18}, 103307 (2006). 

\bibitem{ar2004} I.S. Aranson, private communication, (2004).

\end{thebibliography}
\end{document}